\documentclass[
reprint,
superscriptaddress,
aip,
longbibliography,
showpacs,
jcp,
floatfix
]{revtex4-2}

\usepackage[utf8]{inputenc}

\usepackage{amsmath}
\usepackage{amsfonts}
\usepackage{amssymb,bm}
\usepackage{graphicx}
\usepackage{physics}
\usepackage{upgreek}
\usepackage{xcolor}
\usepackage[outercaption]{sidecap}
\usepackage{orcidlink}
\usepackage{txfonts}

\newcommand{\superL}{\hat{\hat{\mathcal{L}}}}
\newcommand{\superR}{\hat{\hat{\mathcal{R}}}}

\newcommand{\bu}{\mathbf{u}}
\DeclareMathAlphabet{\pazocal}{OMS}{zplm}{m}{n}

\begin{document}

\makeatletter
\let\@fnsymbol\@fnsymbol@latex
\@booleanfalse\altaffilletter@sw
\makeatother
	
\title{Engineering the uncontrollable: Steering noisy spin-correlated radical-pairs with coherent and \emph{in}coherent control}
\author{Farhan T.\ Chowdhury\,\orcidlink{0000-0001-8229-2374}}
\affiliation{Department of Physics and Astronomy, University of Exeter, Exeter EX4 4QL, United Kingdom.}
\affiliation{Living Systems Institute, University of Exeter, Stocker Road, Exeter EX4 4QD, United Kingdom.
}
\affiliation{
Department of Chemistry, The Ohio State University, Columbus, Ohio 43210, United States.}
\author{Luke D.\ Smith\,\orcidlink{0000-0002-6255-2252}}
\affiliation{Department of Physics and Astronomy, University of Exeter, Exeter EX4 4QL, United Kingdom.}
\affiliation{Living Systems Institute, University of Exeter, Stocker Road, Exeter EX4 4QD, United Kingdom.
}
\author{Daniel R.\ Kattnig\,\orcidlink{0000-0003-4236-2627}}
\affiliation{Department of Physics and Astronomy, University of Exeter, Exeter EX4 4QL, United Kingdom.}
\affiliation{Living Systems Institute, University of Exeter, Stocker Road, Exeter EX4 4QD, United Kingdom.
}

\begin{abstract}
The quantum control of spin-correlated radical pairs (SCRPs) holds promise for the targeted manipulation of magnetic field effects, with potential applications ranging from the design of noise-resilient quantum information processors to genetically encodable quantum sensors. However, achieving precise handles over the intricate interplay between coherent electron spin dynamics and incoherent relaxation processes in photoexcited radical-pair reactions requires tractable approaches for numerically obtaining controls for large, complex open quantum systems. Employing techniques relying on full Liouville-space propagators becomes computationally infeasible for large spin systems of realistic complexity. Here, we demonstrate how a control engineering approach based on the Pontryagin Maximum Principle (PMP) can offer a viable alternative by reporting on the successful application of PMP-optimal control to steer the coherent and incoherent spin dynamics of noisy radical pairs. This enables controls for prototypical radical-pair models that exhibit robustness in the face of relevant noise sources and paves the way to incoherent control of radical-pair spin dynamics.

\end{abstract}

\maketitle

Effective means of controlling the dynamics of complex open quantum systems are crucial to the realisation of emerging quantum technologies. Although dissipative effects in the conventional system-bath setup are usually obstacles to surmount for controlling quantum systems, curiously, in some scenarios, they could in principle be harnessed as a potential control knob \cite{kosloff22, brumer18, kurizki2016, pechen12, shapiro1995}. In fact, for spin-correlated radical pairs, which are often implicated to be of potential physiological relevance \cite{gerhards2025rf, rwcont25}, electron spin relaxation and spin decoherence have been shown to boost the magnetometric sensitivity of a cryptochrome chemical compass \cite{kattnig16, luo24}. While this opens the door to the potential incoherent modulation of dissipative degrees of freedom in magnetosensitive radical pairs via optimal quantum control \cite{cont24, prx20q}, exploring this avenue theoretically poses significant computational challenges. In particular, gradient-based approaches for control engineering become numerically intractable for the complete Liouville space description of open-system dynamics for quantum systems of realistic complexity \cite{lewis2025, osqc24, aplQ, magann2024tradeoffs}, thereby severely limiting predictive and interpretive power over experimental realisations. 

In this communication, we demonstrate an approach that circumvents the bottleneck of calculating gradients of Liouville-space propagators (as required, for instance, in approaches like OpenGRAPE \cite{openGR21}) by leveraging a Pontryagin Maximum Principle (PMP) \cite{ansel24, opqc23, Boscain_2021} framework for obtaining both coherent and incoherent controls for photo-induced radical-pair reactions, with its significantly reduced computational overhead allowing for tractable numerical treatments of their full open quantum system dynamics. Starting from the time-dependent density operator $\hat{\rho}(t)$ for the combined electron and nuclear spin system underlying the kinetics of the radical-pair reaction \cite{ivanov17}, we can describe the time evolution under the total spin Hamiltonian in terms of the master equation \cite{Haberkorn1976, Ivanov2010} ($\hbar =1$)

\begin{align}
\dv{}{t}\hat{\rho}(t) = -i[\hat{H}(t), \hat{\rho}(t)] - \frac{k_\mathrm{b}}{2}\{\hat{P}_{\mathrm{S}}, \hat{\rho}(t)\} - k_\mathrm{f}\hat{\rho}(t) + \superR(t) \hat{\rho}(t), \label{eq.master_equation1}
\end{align} 
where the first term accounts for the Liouville--von Neumann evolution, $k_{\mathrm{b}}$ and $k_{\mathrm{f}}$ are reaction rate constants for singlet-state recombination and a spin-independent competing reaction. $\hat{P}_{\mathrm{S}}$ and $\hat{P}_{\mathrm{T}}$ are the projection operators onto the singlet and triplet subspaces, obeying $\hat{P}_{\mathrm{S}}+\hat{P}_{\mathrm{T}} = \hat{I}$, and spin relaxation is represented via the superoperator $\superR(t)$. The latter can, for example, account for quantum noise \cite{horsley25, talukdar24, bremermann1967} arising from fluctuations of interradical interactions \cite{yoshida25, drive22, birad90} and magnetic field fluctuations, referred to as random-field noise, which can be anisotropic, depending on the orientation of radicals in the molecular frame. This form of the master equation has remained a staple for explaining the effects of weak magnetic fields in spin chemistry. While the corresponding choice of recombination term has attracted debate \cite{clausen2014multiple}, the form in Eq.~\eqref{eq.master_equation1} was found to be consistent with experimental results in a carotenoid-porphyrin-fullerene system \cite{Maeda2013}, and has been derived from a microscopic description \cite{fay2018spin}. Singlet-triplet dephasing induced by recombination can also be reflected in $\superR(t)$, highlighting the principal relevance of an open-system treatment. 

\begin{figure}
    \centering
    \includegraphics[width=\linewidth]{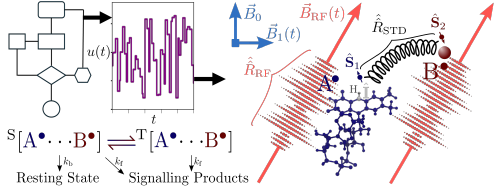}
    \caption{Schematic of our approach, utilising an algorithm based on the Pontryagin Maximum Principle to generate control modulations $u(t)$. These are generated for a system of electron spins $\mathbf{\hat S}_1$ and $\mathbf{\hat S}_2$ of a radical-pair, formed from radicals $\mathrm{A}^{\bullet}$ and $\mathrm{B}^{\bullet}$, which are subject to interradical interactions, hyperfine coupling with nuclear spins $\mathbf{\hat I}$, and Zeeman interaction with a static magnetic field $\vec{B}_{0}$. In addition, the spins can experience relaxation processes due to random field noise $\superR_{\mathrm{RF}}$, arising from magnetic field fluctuations, and singlet-triplet dephasing noise $\superR_{\mathrm{STD}}$, arising, for example, from recombination and fluctuations in the interradical coupling. We consider both coherent control, via a time-dependent field $\vec{B}_{1}(t)$ perpendicular to $\vec{B}_{0}$, and incoherent control over $\superR_{\mathrm{RF}}$ and $\superR_{\mathrm{STD}}$. These controls influence the interconversion between singlet $^\mathrm{S}[\mathrm{A}^{\bullet}\cdot\cdot\cdot \mathrm{B}^{\bullet}]$ and triplet $^\mathrm{T}[\mathrm{A}^{\bullet}\cdot\cdot\cdot \mathrm{B}^{\bullet}]$ spin configurations and subsequent chemical outcomes, with rates $k_{\mathrm{b}}$ and $k_{\mathrm{f}}$, guided by a reaction yield cost function. Shown here is an example $[\mathrm{FADH}^{\bullet} / \mathrm{Z}^{\bullet}]$ radical-pair, with FADH$^{\bullet}$ (blue) and its  most significant anisotropic N$5$ hyperfine coupling highlighted (silver), and a Z$^{\bullet}$-radical devoid of hyperfine couplings represented by a red sphere.}
\end{figure}

Formally, we consider a time-independent spin Hamiltonian comprising the Zeeman interaction of the electron spins with an external magnetic field, hyperfine interactions between electron and nuclei spins, and exchange interactions between electron spins, yielding the prototypical Hamiltonian 
\begin{align}
     \hat H_{0} &= {\hat H_{\rm{Zeeman}}} + {\hat H_{\rm{Hyperfine}}} + {\hat H_{\rm{Exchange}}} \nonumber  \\
 &= \sum_{i}{\vec\omega _i} \cdot {{\mathbf{\hat S}}_i} + \sum_{i}\sum\limits_j {{{{\mathbf{\hat S}}}_i} \cdot {{\mathbf{A}}_{i,j}} \cdot {{{\mathbf{\hat I}}}_{i,j}}} -2 j_\mathrm{ex}\; \mathbf{\hat S}_1 \cdot \mathbf{\hat S}_2 ,
\end{align}
where $\mathbf{\hat S}_i$ denotes the electron spin in the $i$th radical and $\mathbf{\hat I}_{i,j}$ denotes the $j$th nuclear spin coupled to the $i$th electron spin via the hyperfine coupling tensors $\mathbf{A}_{i,j}$. The Larmor precession frequency $\vec{\omega} _i = - \gamma_{i}\mathbf{B}$ is given by the product of the magnetic field $\mathbf{B} = B_{0}(\sin \theta \cos \phi, \sin \theta \sin \phi, \cos \theta)$ and the electron gyromagnetic ratios, with $B_{0}$ denoting the magnitude, and $j_\mathrm{ex}$ being the exchange interaction. In the optimal control setting, we introduce a tunable parameter $\bu(t)$ such that
\begin{align}
    \frac{\mathrm{d}}{\mathrm{d}t}\hat{\rho}(t) = \superL(\bu(t), t) \hat{\rho}(t) = \left(\superL_{0} +\sum_{i=1}^{n}u_{i}(t)\superL_{i}  \right)\hat{\rho}(t),
\end{align}
where the Liouvillian $\superL(\bu(t),t)$ defines the time evolution. $\superL_{0}\hat{\rho}(t) =-i[\hat{H}_{0},\hat{\rho}(t)]-\frac{k_{\mathrm{b}}}{2}\hat{\rho}(t)-k_{\mathrm{f}}\hat{\rho}(t)+\superR \hat{\rho}(t)$ encompasses the drift dynamics, including terms such as coherent evolution under the spin Hamiltonian, recombination dynamics, and spin relaxation as in Eq.~\eqref{eq.master_equation1}. The terms $\superL_{i}$ correspond to control components that couple the system to $\bu(t)$, which may be a coherent contribution to the evolution of the system (coherent control) or enter as tunable dissipators (incoherent control). For convenience, we vectorise the density operator as $\vert \rho(t)\rangle = \mathrm{vec}(\hat{\rho}(t))$, so that the evolution can now be described by
\begin{align}
\dv{}{t}|\rho(t)\rangle=\hat{\mathcal{L}}(\bu(t),t)|\rho(t)\rangle = \left(\hat{\mathcal{L}}_{0} + \sum_{i=1}^{n}u_i(t)\hat{\mathcal{L}}_{i} \right)|\rho(t)\rangle.
\end{align}
To obtain a target output for the control, we define the cost function as
\begin{align}
    G[\bu(t)] = \int_{t_0}^{t_1} I(\vert{\rho}(t)\rangle, \bu(t)) \, \mathrm{d}t,
\end{align}
where the instantaneous cost
\begin{align}
I(|\rho(t)\rangle, \bu(t)) = k_\mathrm{b} \langle P_{\mathrm{S}}|\rho(t)\rangle
\end{align}
corresponds to the singlet recombination flux at time $t$. Thus, $G[\bu(t)]$ corresponds to the recombination yield during the time interval $t_0$ to $t_1$. From this, we define the control objective to be maximisation or minimisation of the recombination yield, i.e. $\max_{\bu(t)}G[\bu(t)]$ or $\min_{\bu(t)}G[\bu(t)] = \max_{\bu(t)}-G[\bu(t)]$, the latter realised by a simple sign flip). To attain a practical scheme for the constrained optimisation problem, we introduce Lagrange multipliers in the form
\begin{align}
L = \int_{t_0}^{t_1} I(|\rho(t)\rangle, \bu(t)) 
+ \langle \lambda(t)|\left( \hat{\mathcal{L}}(\bu(t),t)|\rho(t)\rangle - |\dot{\rho}(t)\rangle \right) \mathrm{d}t , \label{eq:lagrange}
\end{align}
where $\langle \lambda(t)|$ ensures that the evolution equation is obeyed at time $t$. Adding such a constraint for any time yields the formulation above employing an integral over constraints scaled by a Lagrange multiplier function as a generalisation of the usual approach of accommodating multiple constraints.
Performing integration by parts for the last term yields
\begin{align}
\int_{t_0}^{t_1} \langle \lambda(t)|\dot\rho(t)\rangle \,\mathrm{d}t\nonumber=& \langle \lambda(t_1)|\rho(t_1)\rangle - \langle \lambda(t_0)|\rho(t_0)\rangle \\ &- \int_{t_0}^{t_1} \langle \dot{\lambda}(t)|\rho(t)\rangle \, \mathrm{d}t,
\end{align}
which, substituted back into Eq.~\eqref{eq:lagrange}, gives
\begin{align}
    L =& \int_{t_0}^{t_1} \big[I(|\rho(t)\rangle, \bu(t)) + \langle \lambda(t)|\hat{\mathcal{L}}(\bu(t),t)|\rho(t)\rangle+\langle \dot{\lambda}(t)|\rho(t)\rangle\big] \,\mathrm{d}t \nonumber \\ &-\langle \lambda(t_1)|\rho(t_1)+\langle \lambda(t_0)|\rho(t_0)\rangle.
\end{align}
In deriving the first-order condition for an optimum, we assume that a solution has been found, and the Lagrangian is maximised. Then, any perturbation to $\vert \rho(t)\rangle$ or $\bu(t)$ must cause the value of the Lagrangian functional to decline. In particular, the total derivative of $L$ obeys
\begin{align}
\begin{split}
\mathrm{d}L =& \int_{t_0}^{t_1} \Bigl[ I_\bu(|\rho(t)\rangle, \bu(t)) + \langle \lambda(t)|\hat{\mathcal{L}}_\bu(\bu(t),t)|\rho(t)\rangle \Bigr] \, \mathrm{d}\bu(t) \, \mathrm{d}t\\
\quad &+ \int_{t_0}^{t_1} \Bigl[ I_\rho(|\rho(t)\rangle, \bu(t)) + \langle \lambda(t)|\hat{\mathcal{L}}(\bu(t),t) + \langle \dot{\lambda}(t)| \Bigr] \, \vert \mathrm{d}\rho(t) \rangle\, \mathrm{d}t\\
\quad &- \langle \lambda(t_1)|\mathrm{d}\rho(t_1)\rangle + \langle \lambda(t_0)|\mathrm{d}\rho(t_0)\rangle, \label{eq:L_tot_deriv}
\end{split}
\end{align}
where subscripts $\bu$ and $\rho$ denote derivatives, e.g.\ such that $I_\bu(|\rho(t)\rangle, \bu(t)) = \frac{\partial I(|\rho(t)\rangle, \bu(t))}{\partial \bu}$.
For Eq.~\eqref{eq:L_tot_deriv} to equal zero for arbitrary perturbations, the following optimality conditions must necessarily hold:
\begin{subequations}
	\begin{align}
	I_{\bu}(|\rho(t)\rangle, \bu(t)) + \langle\lambda(t) |\hat{\mathcal{L}}_{\bu}(\bu(t),t)|\rho(t)\rangle = 0, \\
	I^{\dagger}_{\rho}(|\rho(t)\rangle, \bu(t)) + \hat{\mathcal{L}}^{\dagger} (\bu(t), t)| \lambda(t)\rangle + |\dot{\lambda}(t)\rangle = 0,
	\end{align}
\end{subequations}
with $|\lambda(t_1)\rangle=0$, as the terminal value of $\rho$ is free (transversality condition for a fixed-horizon problem).

We proceed to define an ancillary function, which we refer to as the control Hamiltonian, as
\begin{align}
H_{\mathrm{c}}( \vert \rho(t) \rangle, \bu(t), \langle\lambda(t) \vert,t) = I(\vert \rho(t) \rangle, \bu(t)) + \langle\lambda(t)|\hat{\mathcal{L}}(\bu(t),t) \vert \rho(t)\rangle .
\end{align}
This Hamiltonian obeys the Pontryagin Maximum Principle (PMP), asserting that the optimal state trajectory \(\rho^*\) and control \(u^*\) along with the corresponding Lagrange multipliers \(\lambda^*\) are the ones that maximise the Hamiltonian \(H_\mathrm{c}\) $\forall$ \(t \in [t_0, t_1]\) for all admissible control inputs \(u \in U\), subject to given constraints, costate dynamics, and necessary boundary conditions.
Indeed, the first-order necessary conditions for a PMP optimum can be generated from the Hamiltonian, specifically,
\begin{widetext}
	\begin{gather}
    \frac{\partial}{\partial \bu} H_{\mathrm{c}}( \vert \rho(t) \rangle, \bu(t), \langle\lambda(t) \vert,t) = I_\bu (\vert \rho(t) \rangle, \bu(t)) + \langle\lambda(t)|\hat{\mathcal{L}}_\bu (\bu(t),t)|\rho(t)\rangle = 0,
    \end{gather}
\end{widetext}
which confirms the maximum principle. Furthermore, 
\begin{align}
\frac{\partial}{\partial \langle\lambda|} H_\mathrm{c}(\vert\rho(t)\rangle, \bu(t), \langle\lambda(t) \vert,t) =& \hat{\mathcal{L}} (\bu(t),t)|\rho(t)\rangle = \vert \dot{\rho}(t)\rangle,
\end{align}
generates the equation of motion, and
\begin{align}
\frac{\partial}{\partial |\rho\rangle} H_{\mathrm{c}}( \vert \rho(t) \rangle, \bu(t), \langle\lambda(t) \vert,t)=& I_\rho (\vert \rho(t) \rangle, \bu(t)) + \langle\lambda(t) \vert \hat{\mathcal{L}}_{\rho}(\bu(t), t) \nonumber \\ =& -\langle\dot{\lambda}(t)|, \label{eq:costate}
\end{align}
the corresponding costate equations. The PMP stipulates that it is necessary for any optimal control along with the optimal state trajectory to solve the so-called Hamiltonian system, which is a two-point boundary value problem, plus a maximum condition of the control Hamiltonian \cite{pontryagin1962mathematical}.

To practically optimise the reaction yield of the radical pair (or the reaction yield difference for two chosen directions of the magnetic field), we follow an iterative approach to optimise the piecewise constant control amplitudes. In every step, the current values of the controls are used to solve the forward and co-state equations, storing the respective states corresponding to the individual control boundaries. For the forward pass, we step the system forward in time by formally applying the one-control-step propagator to the state, whilst avoiding its explicit construction using the approach of Al-Mohy and Higham \cite{Al-mohy_2011, Higham_2010}, which relies on a truncated Taylor series approximation to the exponential directly applied to the state. For the co-states, we propagate the system back in time using an explicit Runge-Kutta method of order 8 for every control step \cite{Hairer_1993_multistep}. Attempts to use the analytical solution of Eq.~\eqref{eq:costate} together with the exponential propagator as for the forward pass were abandoned, as the required linear solution step to apply the inverse Liouvillian turned out to be non-competitively slow. With the states and co-states available, we then calculate the gradient of the Hamiltonian with respect to the controls, which is used as the search direction in a subsequent line search optimising the reaction yield (or the reaction yield difference). We implemented an elementary line search approach, with the learning rate initially set to allow a maximal change in the individual control amplitudes of $0.1$. If the resulting update of controls does not provide an increase in the optimisation criterion, the learning rate is halved in every unsuccessful step, and the search iterated. The learning rate is typically retained for the subsequent gradient optimisation step. However, every 10th step we have reset the learning rate to its respective maximum, which in tests has shown a marked speedup of the optimisation. After completion of the line search, a new search direction is determined from the gradient of the optimisation Hamiltonian, provided that at this stage the step length, change in gradient with respect to the previous iteration, or gradient norm have not fallen below a user-set tolerance, and the maximal number of iterations has not been exceeded. For the problems considered here, a few tens of iterations typically sufficed for the reaction yield to change by less than the $10^{-4}$th part. The algorithm was implemented in Python based on NumPy \cite{harris2020_numpy} and the SciPy \cite{Virtanen_2020_scipy} functions \texttt{expm\_multiply} and \texttt{solve\_ivp} (with DOP853 method). 

\begin{figure*}
    \centering
    \includegraphics[width=0.9\linewidth]{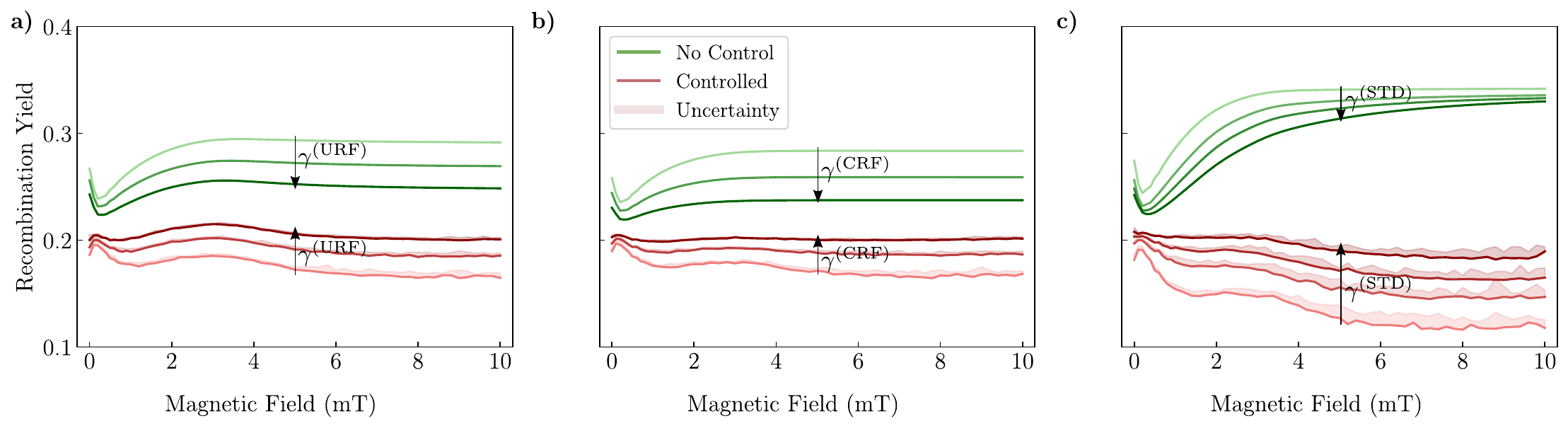}
    \caption{\label{Fig:maeda} Minimising the singlet recombination yield via coherent control of the prototypical radical-pair system from \cite{mae} (described in the text) in the presence of a) uncorrelated random-field noise (URF), b) correlated random-field noise (CRF), and c) singlet–triplet dephasing noise (STD), respectively. Solid curves represent the best-performing control field from 9 replications, while shaded bands indicate the region between the maximum and the 80th percentile of the recombination yield for the various control attempts. Noise rates used were $\gamma^{(\mathrm{URF})}=\gamma^{(\mathrm{CRF})}=0.5~\mu\mathrm{s}^{-1}$, $1~\mu\mathrm{s}^{-1}$, and $2~\mu\mathrm{s}^{-1}$ for URF and CRF; and $\gamma^{(\mathrm{STD})}=1~\mu\mathrm{s}^{-1}$, $5~\mu\mathrm{s}^{-1}$, $10~\mu\mathrm{s}^{-1}$, and $20~\mu\mathrm{s}^{-1}$ for STD, with the arrows pointing in the direction of increasing noise rate.}
\end{figure*}

We first demonstrate the success of our PMP scheme in minimising the singlet recombination yield of a radical pair, that is, we consider $\min_{\bu(t)}G[\bu(t)]$. We focus on a generalisation of the prototypical radical pair studied by Masuzawa et al.\ \cite{mae}, which we extend to include asymmetric recombination and various forms of spin relaxation. Our complete Liouville space description allows coherent controls to be derived while incorporating incoherent processes, such as recombination and spin relaxation. While we have considered asymmetric recombination through our GRAPE approach in our prior work \cite{cont24}, no previous approach has considered open systems involving spin relaxation. Specifically, here we consider a 7-spin system comprising two electron spins each coupled to nuclear spins, one to 3 and the other to 2, via isotropic hyperfine interactions ($0.2$, $0.5$ \& $1\mskip3mu$mT, and $0.2$ \& $0.3\mskip3mu$mT), including an exchange coupling ($j_{\mathrm{ex}}/(2\pi) = 1\mskip3mu$MHz), and undergoing spin-selective recombination in the singlet state with rate constant $k_\mathrm{b} = 1 \mskip3mu\mu$s$^{-1}$ or spin-independent escape with rate constant $k_\mathrm{f} = 1 \mskip3mu\mu$s$^{-1}$, such that the effective singlet and triplet decay rates are $k_{\rm S} = k_\mathrm{b} + k_\mathrm{f}$ and $k_{\rm T} = k_\mathrm{f}$, respectively. As in \cite{cont24}, we apply a control field $\omega_{1}(\hat{\mathbf{S}}_{1,x} + \hat{\mathbf{S}}_{2,x}$) perpendicular to a static biasing field, where $\omega_{1}$ is the maximal Rabi frequency of the control field, i.e.\ $|u| < 1$. We used 1000 piecewise constant controls that spanned $1\mskip3mu$ns each, i.e., we controlled the first $1\mskip3mu\mu$s after radical pair generation. The optimisation used 25 gradient evaluations, performing a line search to minimise the recombination yield in every step; the yield typically levelled off towards its final value after 10 iterations. The controls were initialised from a normally distributed random variable with a standard deviation of 0.1 centred on 0. The final yield reported in Fig.~\ref{Fig:maeda} was evaluated with $t_1$ = 10 $\mu$s. We consider spin relaxation arising from singlet-triplet dephasing (STD), uncorrelated random-field noise (URF), and correlated random-field noise (CRF). Fluctuations in interradical interactions can lead to STD, while URF and CRF can arise from fluctuations in interactions coupling to the electron spins or the applied magnetic field. In terms of Lindblad dissipators, which act with collapse operator $\hat{A}$ on state $\hat{\rho}$ as $\mathcal{D}[\hat{A}]\hat{\rho} = \hat{A}\hat{\rho}\hat{A}^{\dagger} - \frac{1}{2}\{\hat{A}^{\dagger}\hat{A}, \hat\rho\}$ with $\{\cdot, \cdot\}$ denoting the anticommutator, the relaxation superoperators are given by
\begin{align}
    \superR_{\mathrm{STD}}\hat{\rho} =& \gamma^{(\mathrm{STD})} \mathcal{D}[\hat{P}_{\mathrm{S}}]\hat{\rho}, \\
    \superR_{\mathrm{URF}}\hat{\rho} =& \gamma^{(\mathrm{URF})} \sum_{i=1,2}\sum_{\alpha=x,y,z} \mathcal{D}[\hat{S}_{i,\alpha}]\hat{\rho}, \\
    \superR_{\mathrm{CRF}}\hat{\rho} =& \gamma^{(\mathrm{CRF})} \sum_{\alpha=x,y,z}\mathcal{D}[\hat{S}_{1,\alpha}+\hat{S}_{2,\alpha}] \hat{\rho},
\end{align}
with noise rates $\gamma^{(\mathrm{STD})}$, $\gamma^{(\mathrm{URF})}$, and $\gamma^{(\mathrm{CRF})}$, respectively, where the latter two noise rates have been assumed to be isotropic. 

\begin{figure*}
    \centering
    \includegraphics[width=1\linewidth]{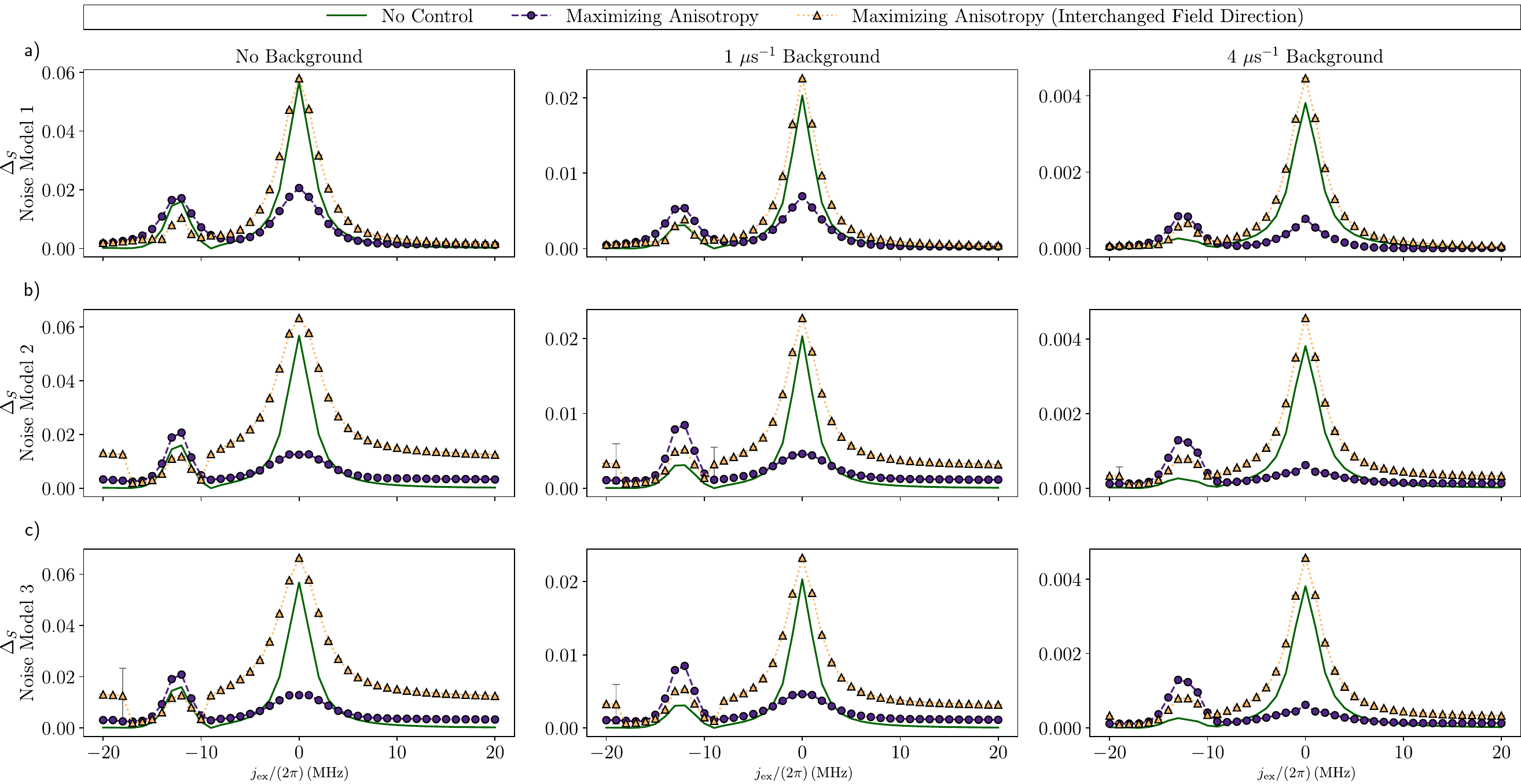}
    \caption{Anisotropic magnetic field effects optimised by application of controlled directional noise for various noise scenarios as a function of the exchange interaction $j_{\mathrm{ex}}$. For a simple one-nitrogen radical-pair, the spread of the recombination yield as a result of reorienting the geomagnetic field direction has been optimised by application of timed directional noise in the absence (left) and presence of uncorrelated random field noise of rate 1$~\mu\mathrm{s}^{-1}$(middle) and 4$~\mu\mathrm{s}^{-1}$ (right). We optimized the added noise assuming a) correlated pairwise control (CPC) noise of both electrons, b) uncorrelated pairwise control (UPC) noise, and c) uncorrelated independent control (UIC) noise, as described in Table \ref{tab:noise_models}. The control noise amplitude was capped at $6\mskip3mu\mu$s$^{-1}$ and we controlled the initial $2\mskip3mu\mu$s using $2000$ piecewise constant control steps ($t_1 = 2\mskip3mu\mu$s, $t_0 = 0\mskip3mu\mu$s).}
    \label{noise}
\end{figure*}

As seen in Fig.~\ref{Fig:maeda}, our PMP scheme allows one to derive coherent controls in the presence of spin relaxation for a spin system of a size that has been deemed impractically large even for purely coherent GRAPE-based approaches that neglect spin relaxation \cite{cont24, mae}. However, this is made possible here, as once the state and costate have been constructed by a forward and backward time propagation, gradients of the control Hamiltonian are trivial to evaluate. This contrasts with approaches which require the gradient of the exponential propagators \cite{openGR21}, which are dense and, when applied to a Liouville space formulation, prohibitively costly to evaluate even for spin systems of moderate complexity. Fig.~\ref{Fig:maeda} shows how the magnetic field sensitivity is generally reduced by spin relaxation. However, the system remains susceptible to coherent controls and, remarkably, can be controlled to a recombination yield not reachable by the application of a static magnetic field of any amplitude regardless of noise amplitude. Furthermore, it is evident that a more significant change in recombination yield can be achieved in the presence of STD, for which large noise rates of $20\,\mu\mathrm{s^{-1}}$ can be accommodated with little effect on the MFE amplitude and the existence of the low-field effect and with only moderate broadening. However, the controllability appears to be more strongly affected, in particular in higher magnetic fields, where the spread of recombination yields for different noise rates is clearly seen to exceed the corresponding spread in the uncontrolled case (see Fig.~\ref{Fig:maeda}C). It is furthermore noteworthy that with strong noise the controlled recombination yield loses its sensitivity to static magnetic field strength, i.e.\ the strength of the biasing field becomes less critical, while for low noise larger biasing fields provide the best contrast with respect to the uncontrolled case. 

\begin{table*}
\caption{Explicit relaxation superoperators formed of Lindblad dissipators $\mathcal{D}[\cdot]$ for each noise model, with spin operators $\hat{\mathbf{S}}_{i}$ acting on electron spins $i \in 1,2$. While all models consider separate equatorial/axial noise control, correlated noise corresponds to the scenario where the dissipators act across both spins, i.e.\ $\mathcal{D}[\hat{S}_{1,j} + \hat{S}_{2\beta}]$, while uncorrelated noise corresponds to dissipators that act on individual electron spin operators $\mathcal{D}[\hat{S}_{i,\alpha}]$ for $\alpha \in x,y,z$. Pairwise control refers to simultaneous tuning of noise acting on both electron spins, while independent control tunes noise on electron spins separately, where $\gamma$s encapsulate the noise strength.}
\label{tab:noise_models}
\renewcommand{\arraystretch}{1.5}
\begin{ruledtabular}
\begin{tabular}{ll}
 \textbf{Noise Model for Incoherent Control} & \textbf{Incoherent Control Relaxation Superoperators} \\[3pt]
\hline

\text{1: Correlated Pairwise Control Noise (CPC)} &
$\superR_{\mathrm{CPC}} = \gamma^{(1)}\left( \mathcal{D}[\hat{S}_{1,x} + \hat{S}_{2,x}] + \mathcal{D}[\hat{S}_{1,y} + \hat{S}_{2,y}] \right)
+ \gamma^{(2)}\mathcal{D}[\hat{S}_{1,z} + \hat{S}_{2,z}]$ \\[6pt]

\text{2: Uncorrelated Pairwise Control Noise (UPC)} &
$\superR_{\mathrm{UPC}} = \gamma^{(1)}\left( \mathcal{D}[\hat{S}_{1,x}] + \mathcal{D}[\hat{S}_{2,x}] + \mathcal{D}[\hat{S}_{1,y}] + \mathcal{D}[\hat{S}_{2,y}] \right)
+ \gamma^{(2)}\left( \mathcal{D}[\hat{S}_{1,z}] + \mathcal{D}[\hat{S}_{2,z}] \right)$ \\[6pt]

\text{3: Uncorrelated Independent Control Noise (UIC)} &
$
\begin{aligned}
\superR_{\mathrm{UIC}} \, =\ & \gamma^{(1)}\left( \mathcal{D}[\hat{S}_{1,x}] + \mathcal{D}[\hat{S}_{1,y}] \right)
+ \gamma^{(2)}\left( \mathcal{D}[\hat{S}_{2,x}] + \mathcal{D}[\hat{S}_{2,y}] \right) \\ &+ \gamma^{(3)}\mathcal{D}[\hat{S}_{1,z}]
+ \gamma^{(4)}\mathcal{D}[\hat{S}_{2,z}]
\end{aligned}$
\end{tabular}
\end{ruledtabular}
\end{table*}

Given the resilience of controlling the system to the presence of noise, a further question is raised; could controlling noise also prove effective to enhance magnetosensitive traits, such as the optimality of a radical-pair-based compass sense \cite{contQFIdr25, opt24, probing24} in migrating organisms? We thus devote our attention to modulating incoherent magnetic noise parameters with the goal of optimising the directional magnetic field sensitivity for a simple radical pair model of the proposed chemical compass mechanism implicated in migratory navigation, the $[\mathrm{FADH}^{\bullet}/ \mathrm{Z}^{\bullet}]$ model. Our simple model incorporates the most dominant N$5$ axial hyperfine coupling tensor of a flavin radical (as found in the putative magnetoreceptive protein cryptochrome; we use an axial hyperfine tensor with $A_{xx}=A_{yy}=-2.6\,$MHz and $A_{zz}=49.2\,$MHz) in combination with a radical devoid of hyperfine interactions (representative of superoxide \cite{denton2024, deviers2024, Lee2014, Ritz2009}, a potential candidate for the Z radical). This particular well-studied system \cite{Lee2014, Cai2012, Gauger2011}, which has been termed the reference-probe radical pair \cite{Lee2014, Procopio2020, Ritz2010}, boasts increased magnetometric sensitivity compared to systems that have strong hyperfine interactions in both radicals. Spin relaxation and noise remain significant in these systems, and controlling these processes raises an intriguing prospect. It is furthermore tempting to posit that for a highly evolutionarily optimised trait, such as magnetoreception, shaped noises could be induced by the inherently non-Markovian characteristics of the environment \cite{ding2025, dodin24, petruhanov2023incoherent, drive22}, potentially in combination with or bolstered by signatures of chirality \cite{fay25, chir_gates25, smith2025_ciss_zeno, Foo2025, kermiche22}, electron hopping \cite{Ilott2025}, or the quantum zeno effect \cite{denton2024}. 

\begin{figure*}
    \centering
    \includegraphics[width=0.9\linewidth]{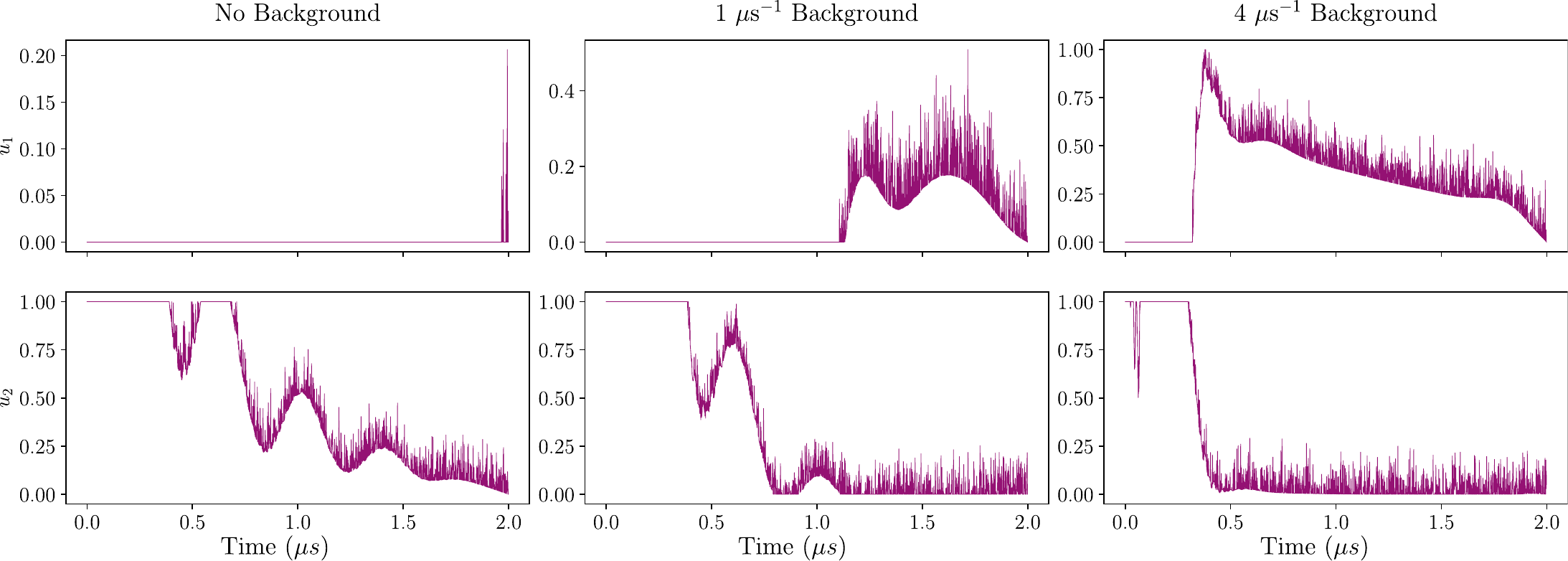}
    \caption{\label{pulses}Modulations without (left) and with the presence of URF noise of rate 1$~\mu\mathrm{s}^{-1}$ (middle) and 4$~\mu\mathrm{s}^{-1}$ (right) with \( j_{\mathrm{ex}}/(2\pi) = 2\mskip3mu\text{MHz} \) in $[\mathrm{FADH}^{\bullet}/ \mathrm{Z}^{\bullet}]$ for incoherent control using uncorrelated pairwise control (UPC) noise where $u_1$ modulates the equatorial component of $\superR_{\mathrm{UPC}}$ and $u_2$ modulates its axial component with control amplitudes $u_1 , u_{2} < 1$, the maximal amplitudes corresponding to a noise rate of $6\mskip3mu\mu$s$^{-1}$. Noticeably, the first $0.5\,\mu\mathrm{s}$ appears as a crucial point to exert control, likely reflecting the period of highest coherent interconversion and significant radical-pair population.}
\end{figure*}

Building on earlier investigations showing that spin relaxation can enhance a cryptochrome-based magnetic compass sensor \cite{kattnig16, luo24}, we first consider correlated equatorial ($x$,$y$) and axial ($z$) noise (see Table.\ \ref{tab:noise_models}; CPC), followed by uncorrelated equatorial and axial noise (UPC), and finally fully separated uncorrelated noise, where each noise rate is tuned independently for each spin and equatorial and axial directions (UIC).  We describe each of the three noise models used in incoherent control in terms of relaxation superoperators $\superR_{\mathrm{CPC}}$, $\superR_{\mathrm{UPC}}$, and  $\superR_{\mathrm{UIC}}$, formed of the constituent Lindblad dissipators in Table.~\ref{tab:noise_models}. The anisotropic magnetic field effect was quantified through the difference of the recombination yield,
\begin{align}
\Delta_{S}[\bu(t)] = G[\bu(t);\Omega_{z}] - G[\bu(t);\Omega_{x}],
\end{align}
when the magnetic field was applied along the $z$- and the $x$-direction (orientations $\Omega_{z}$ and $\Omega_{x}$), reflecting the directional yield contrast. Although in general the maximal yield contrast is evaluated sampling many directions $\Omega$, of the magnetic field, here the restriction to two orientations is reasonable due to the symmetry of the system. 
In Fig.~\ref{noise}, we consider three scenarios as a function of the exchange coupling strength, $j_{\mathrm{ex}}$, and both in the presence and in the absence of background URF noise of rates 1$~\mu\mathrm{s}^{-1}$ and 4$~\mu\mathrm{s}^{-1}$. Namely, these are no control (\mbox{$\Delta_{S}[0]=G[0;\Omega_{z}]-G[0;\Omega_{x}]$}), maximising the difference via a shared control ($\max_{\bu(t)}\Delta_{S}[\bu(t)]=\max_{\bu(t)}(G[\bu(t);\Omega_{z}]-G[\bu(t);\Omega_{x}])$) and maximising with interchanged field directions ($\max_{\bu(t)}\Delta_{S}[\bu(t)]=\max_{\bu(t)}(G[\bu(t);\Omega_{x}]-G[\bu(t);\Omega_{z}])$), plotting these as a green solid line, violet circles, and yellow triangles, respectively; with the curves reporting the best results obtained from 10 replications each involving 500 optimisation steps. The control dissipation channels were bounded by $\gamma_{\max} = 6\ \mu\text{s}^{-1}$, that is, larger than the isotropic background rate, ensuring that optimised control fields can dominate over uncontrolled relaxation while remaining within a physically plausible amplitude range. The optimisation is susceptible to occasionally getting stuck in local minima \cite{wiedmann25, cerezo25, glass25, beato24, birtea22, pulse21, koch16}, which is evident for the small number of cases for which anisotropy of the no-control case exceeds that for case including controlled directional noise. For all noise models, we observe that improvements over the case of no control are possible. In particular, uncorrelated pairwise noise control (UPC) produces the largest enhancements to $\Delta_{S}$, extending to large $j_{\mathrm{ex}}$ values, with no clear change if each spin is independently controlled. It is apparent that maximising the anisotropy with interchanged field directions is generally more effective, except in the approximate range of $-15\leq J \leq -10\,$MHz. With respect to background noise, the enhancement is mostly preserved relative to the case of no control, yet there is a noticeable decrease in $\Delta_{S}$ as the noise rate increases above $1\,\mu\mathrm{s}^{-1}$. It is remarkable to see that some degree of modularity over the underlying excited-state spin dynamics is, in principle, possible with solely a handle over incoherent degrees of freedom. In Fig.~\ref{pulses}, we show the control amplitudes for incoherent control over UPC noise, from which it is observed that the optimisation adjusts noise amplitudes gradually, exhibiting a smooth modulation profile, to guide the open-system dynamics (this is different from coherent controls, which typically exhibit fast transients with marked high-frequency components). Incoherent control offers a potentially simple scheme and clear temporal turnover in which significant axial noise is applied early followed by equatorial noise, which is enacted more significantly and swiftly dependent on the background noise rate, potentially pointing to a greater robustness of equatorial noise in strongly decohered regimes. 

PMP-optimal control has been shown to be facilitative across a wide spread of emerging quantum technologies \cite{lee25, Fressecolson_2025, Riosmonje_2025, Dehagani_2022, sels21, prx17}, but previous efforts \cite{slotine25, abdulla2024bang} to extend PMP-based control approaches to treat radical-pair systems were limited in their applicability to only the Schr\"odinger evolution description of quantum dynamics. Beyond precluding incoherent controls, it is unclear if the coherent controls obtained for radical pairs with existing approaches for idealised closed-system settings would apply effectively in systems of realistic complexity with unavoidable system-bath couplings, spin relaxation, and experimental imperfections \cite{kuprov25, tiersch12}. Here, the PMP approach excels by making possible the engineering of coherent controls in complex open systems \emph{and} opening up the field to incoherent control. Recent work \cite{contQFIdr25} has suggested how driving features \cite{drive22} arising from environment-assisted quantum dynamics \cite{noise25, bec25, noise24, denton2024, Dodin_2021} can push magnetosensing in radical-pair systems closer to approaching quantum limits of metrological precision \cite{opt24, probing24}. Our approach, in extending the PMP approach to treat the Liouville space description of radical-pair excited-state dynamics, offers the potential not only to simulate behaviour for systems of realistic complexity but also to engineer the environment through tuning magnetic noise degrees of freedom, potentially creating a route for noise to be utilised as a resource in addition or synergy with coherent processes \cite{n3fv25, smith2022observations}. It would be of interest for future studies to consider whether a further reduction in computational costs is viable through synergies with wavefunction Monte-Carlo and stochastic Schr\"{o}dinger equation based approaches \cite{Fay_2021, Keens_2020, Smith_2019}, which avoid full density matrix treatments \cite{kappen25, Keijzer_2025, tutorialOQS}. 

In practice, our degree of control over quantum systems is limited, with precisely fine-tuned time-dependent driving or arbitrary entangling unitaries with auxiliary systems often unattainable for even engineered systems \cite{dunlop25c}, let alone for non-equilibrium biochemical processes. Nevertheless, by testing several noise models, we have revealed that it is in principle possible to gain an enhancement in magnetic field sensitivity of spin systems underlying model radical pairs by harnessing phenomenologically motivated noise channels through incoherent control. Although coherent control remains the more effective, as one would expect, there is scope to harness more tailored noise processes, along with broader orientation sweeps, to produce significant enhancements via engineered incoherent modulation \cite{dey25, koch25, mcCaul25, prx20, budker07}. Given the smoother temporal profile and clear scheme for response to an increase in background noise, as seen in our radical-pair incoherent controls, they may in practice be more easily achievable. Experimental realisations that extend control and bath engineering techniques to complex biochemically relevant SCRPs can allow us to glean subtle insights into the interplay of noisy quantum dynamics and information processing \cite{rwcont25, kurizki25} at the boundaries of the quantum-classical transition. This could in turn inform design principles for both noise-resistant molecular quantum information processing architectures and genetically encoded chemical sensors operating near quantum limits \cite{v1b25, feder25, franco25, harvey21, lee11}. In a physiological context, such precise modularity could further open the possibility of mastering metabolic functions through the control of magnetoresponsive pathways \cite{aiello25, aiello24weak}. 

\section*{Acknowledgment}
\noindent F.T.C.\ acknowledges discussions with Dr.\ Joe Dunlop, Prof.\ Oleksandr Kyriienko, Prof.\ David Wineland and Dr.\ Jonas Glatthard. This work was supported by the Office of Naval Research (ONR Award Number N62909-21-1-2018) and the Engineering and Physical Sciences Research Council (EP/X027376/1). The authors acknowledge the use of the University of Exeter High-Performance Computing Facility. For the purpose of open access, a Creative Commons Attribution (CC BY) licence has been applied to any Author Accepted Manuscript version arising from this submission.

\section*{Conflict of Interest}
The authors have no conflicts of interest to disclose.

\section*{References}
%

\end{document}